\documentclass[prl, amsmath, twocolumn,amssymb]{revtex4}
\usepackage{graphics,hyperref}

\newcommand{\ket}[1]{|#1\rangle}
\begin{document}
\date{\today} \title{Quantum stochastic walks: A generalization of classical random walks and quantum
  walks} 

\author{C\'{e}sar A. Rodr\'{i}guez-Rosario} \email[email:
]{rodriguez@chemistry.harvard.edu} \author{James D. Whitfield}
\email[email: ]{whitfield@chemistry.harvard.edu} \author{Al\'{a}n
  Aspuru-Guzik} \email[email: ]{aspuru@chemistry.harvard.edu}
\affiliation{ Department of Chemistry and Chemical Biology, Harvard
  University, 12 Oxford St., Cambridge, MA 02138 }

\begin{abstract}
  We introduce the quantum stochastic walk (QSW), which determines the
  evolution of generalized quantum mechanical walk on a graph that
  obeys a quantum stochastic equation of motion. Using an axiomatic
  approach, we specify the rules for all possible quantum, classical
  and quantum-stochastic transitions from a vertex as defined by its
  connectivity. We show how the family of possible QSWs encompasses
  both the classical random walk (CRW) and the quantum walk (QW) as
  special cases, but also includes more general probability
  distributions. As an example, we study the QSW on a line, the QW
  to CRW transition and transitions to genearlized QSWs that go beyond
  the CRW and QW. QSWs provide a new framework to the study of  quantum algorithms as well as of quantum
  walks with environmental effects.
\end{abstract}


\maketitle 

Many classical algorithms, such as most Markov-chain Monte Carlo
algorithms, are based on classical random walks (CRW), a probabilistic motion through the vertices of a graph. The quantum walk
(QW) model is a unitary analogue of the CRW that is generally
used to study and develop quantum algorithms
\cite{Farhi98a,Venegas08a,Ambainis08a}. The quantum mechanical nature
of the QW yields different distributions for the position of the
walker, as a QW allows for superposition and interference effects
\cite{Aharonov03a}. Algorithms based on QWs exhibit an exponential
speedup over their classical counterparts have been
developed~\cite{childs01,Watrous01a,Childs03a}. QWs have inspired the
development of an intuitive approach to quantum algorithm design \cite{Shenvi03a}, some based
on scattering theory \cite{FarhiGoldstone2008}. They have recently been
shown to be capable of performing universal quantum
computation~\cite{Childs09a}.

The transition from the QW into the classical regime has been studied
by introducing decoherence to specific models of the discrete-time QW
\cite{Brun03a,Kendon07a,Romanelli05a, Love05a}.  Decoherence has also been been
studied as non-unitary effects on continuous-time QW in the context of
quantum transport, such as environmentally-assisted energy transfer in
photosynthetic complexes \cite{Rebentrost08a,Mohseni08a, Plenio08a,Caruso09a,Rebentrost09a}
and state transfer in superconducting qubits
\cite{Strauch08a,Strauch09a}. For the purposes of experimental
implementation, the vertices of the graph in a walk can be implemented
using a qubit per vertex (an inefficient or unary mapping) or by
employing a quantum state per vertex (the binary or
efficient mapping). The choice of mapping impacts the simulation
efficiency and their robustness under decoherence \cite{Hines07a,
  Drezgic08a,Strauch09b}. The previous proposed approaches for
exploring decoherence in quantum walks have {\sl added}
environmental-effects to a QW based on computational or physical
models such as pure dephasing \cite{Rebentrost09a} but have not
considered walks where the environmental effects are constructed
axiomatically from the underlying graph.

\begin{figure}[htb]
\begin{center}
\resizebox{.5\columnwidth}{.37\columnwidth}{\includegraphics{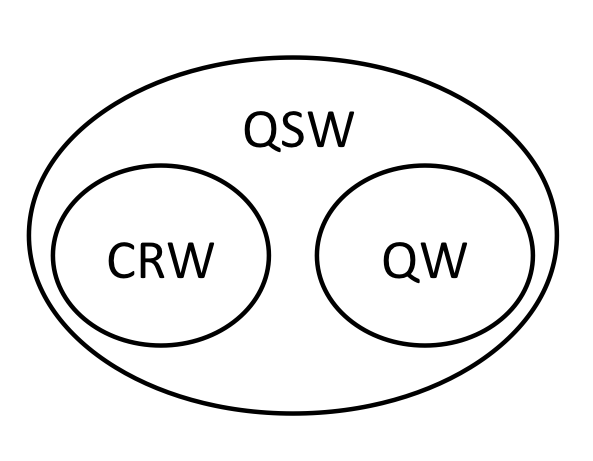}}
\end{center}  
\caption{The quantum stochastic walk (QSW) is a quantum stochastic
  process defined by a set of axioms that describe the connectivity of
  an underlying graph. The family of possible QSWs encompasses the
  classical random walk (CRW) and the quantum walk (QW) as limiting
  cases. QSWs can be used to generate walks that continuously
  interpolate between QW and CRW, and therefore can be employed to
  analyze algorithmic classical-to-quantum transitions.  The family of
  QSWs also describes more general quantum stochastic processes with a
  behavior that cannot be recovered by the QW and CRW approaches.}
\label{venn}
\end{figure}

In this work, we define the quantum stochastic walk (QSW) using a set
of axioms that incorporate unitary and non-unitary effects. A CRW is a
type of classical stochastic processes. From the
point of view of the theory of open quantum systems, the generalization of a classical stochastic process to the quantum
regime is known to be a quantum stochastic process
\cite{Sudarshan61a,Kossakowski72a,Lindblad76a,
  Gorini76a,Rodriguez08,Mohseni08a} which is the most general type of evolution of a density matrix, not simply the Hamiltonian
process proposed by the QW approach. The main goal of this paper is to
introduce a set of axioms that allow for the construction of a quantum
stochastic process constrained by a graph. We call all the
walks that follow these axioms QSWs. We will show that the family of
QSWs includes both the CRW and the QW as limiting cases. The QSW can
yield new distributions that are not found either in the CRW or the QW. The connection
between the three types of walks discussed in this manuscript is
summarized in Fig. \ref{venn}.
For clarity, we focus on continuous-time walks, but also sketch the
corresponding procedure for the discrete-time walks. The QSW provides
the fundamental tools to study the quantum-to-classical transition of
walks, as well as new methods of control by the application of
non-unitary operations on a quantum system.

The CRW describes probabilistic motion over a graph. The dynamics of
the probability distribution function is given by the master equation
for a classical stochastic process,
\begin{equation}
\label{cw}
\frac{d}{dt} p_a=\sum_b M_{b}^a\, p_b,
\end{equation}
where the vector element $p_b$ is the probability of being found at
the vertex $b$ of the graph.  The matrix $M$ is the generator for the
evolution; its structure is constrained by axioms derived from the
connectivity of the graph. For example, if vertices $a$ and $b$ are
connected, $M_{b}^a =-\gamma$, if they are not, $M_{b}^a =0$, and
$M_a^a=d_a\gamma$ where $d_a$ is the degree of vertex
$a$.

In analogy to the CRW, the QW \cite{Farhi98a} has been defined so
that the probability vector element $p_a$ is replaced with $\langle a
\vert \psi \rangle$, which evolves according to the Schr\"odinger
equation,
\begin{equation}
\label{qw}
\frac{d}{dt} \langle  a \vert \psi \rangle=-i\sum_b  \langle a \vert H \vert b \rangle \langle b \vert \psi \rangle,
\end{equation}
where $H$ is a Hamiltonian to be defined based on axioms coming from
the graph.  A choice of this definition is $\langle a \vert H
\vert b \rangle = M_{b}^a$. This unitary evolution effectively rotates
populations into coherences and back \footnote{The Hermitian form of
  the Hamiltonian eliminates directed graphs from consideration,
  motivating work on pseudo-Hermitian
  Hamiltonians~\cite{Salimi09a}.}. The QW fails to completely capture
the stochastic nature of CRW.  The random aspect of the QW comes
solely from the probabilistic nature of measurements performed on the
wavefunction.

Since a classical stochastic process can be generalized to the quantum
regime by means of a quantum stochastic process, a CRW should be
generalized to a QSW derived from the graph \footnote{Just
like a classical stochastic process includes, but is not limited, to
bistochastic processes, a quantum stochastic processes includes but is
not limited to unitary processes \cite{Sudarshan61a}.}.  For the
generalization, we identify the probability vector with elements $p_a$
with a density matrix with elements $\rho_{a\alpha}$, and generalize the
evolution to a quantum stochastic process, $\frac{d}{dt}
\rho=\mathcal{M}\big[\rho\big]$, where $\mathcal{M}$ is a
superoperator \cite{Sudarshan61a,Kossakowski72a, Gorini76a}. To make
this evolution look similar to Eq. (\ref{cw}), we write the density
matrix in terms of its indices,
$\rho=\sum_{a,\alpha}\rho_{a\alpha}\vert a \rangle \langle \alpha
\vert$, and the quantum stochastic master equation becomes,
\begin{equation}
\label{qsw}
\frac{d}{dt} \rho_{a\alpha}=\sum_{b,\beta}\mathcal{M}^{a\alpha}_{b\beta}\;\rho_{b\beta},
\end{equation}
with the tensor $\mathcal{M}^{a\alpha}_{b\beta}= \langle a \vert
\,\mathcal{M}\big[\; \vert b \rangle \langle \beta \vert\; \big]\,
\vert \alpha \rangle$. This correspondence was pointed out by Mohseni
\emph{et al.} in the context of energy transfer \cite{Mohseni08a}.

For a quantum stochastic process to be related to a walk, the
superoperator $\mathcal{M}$ must reflect the 
graph. The connectivity of the vertices will impose conditions on the
transition rates of $\mathcal{M}$. Since the quantum stochastic
process is more general than both the classical stochastic process and
the Schr\"odinger equation, the correspondence of the connectivity of
the graph to the rules imposed on $\mathcal{M}$ should include and go
beyond the connectivity axioms for each of those. For a vertex $m$
connected to vertices that include vertex $n$, we define a processes
$\vert m \rangle \leftrightarrows \vert n \rangle$ which occurs at
some rate that can evolve $\vert m \rangle$ to and from $ \vert n
\rangle$.  Transition rates for vertices that are not connected are
defined to be zero. We employ these connectivity rules as the main
principle for defining valid QSWs from a given
graph. To further explore the connection from the QSW to the CRW and QW as well as
more general behaviors, we discuss the different limiting cases.

For the classical case, the allowed transitions come from incoherent
population hopping of the form $\vert m \rangle \langle m \vert
\leftrightarrows \vert n \rangle \langle n \vert,$ and, for
completeness, $\vert m \rangle \langle m \vert \leftrightarrows \vert
m \rangle \langle m \vert$. These conditions constrain $\mathcal{M}$
to operate only on the diagonal elements of $\rho$, like
Eq. (\ref{cw}). In other words, the QSW transition tensor must have
the property
$\sum_{\alpha,\beta}\delta_{a\alpha}\,\mathcal{M}_{b\beta}^{a\alpha}\,\delta_{b\beta}=
M_b^a$.

The QSW should also recover the QW, where evolution among vertices
happens through coherences developed by a Hamiltonian. These
transitions include terms of the form, $\vert m \rangle \langle m
\vert \leftrightarrows \vert m\rangle \langle n \vert$, that
exchange populations to coherences with the connected vertices. For
completeness, we also consider transitions of the form $\vert m
\rangle \langle n \vert \leftrightarrows \vert m\rangle \langle n
\vert$. If $m$ is also connected to another vertex $l$, additional
transitions can happen. One transition exchanges populations into
coherences among the two vertices connected to $m$, $\vert m \rangle
\langle m \vert \leftrightarrows \vert l\rangle \langle n \vert$; the
other exchanges coherences between $m$ and a connected vertex into
coherence among two vertices connected to $m$, $\vert m \rangle
\langle n \vert \leftrightarrows \vert l\rangle \langle n
\vert$. These conditions allow for the recovery of Eq. (\ref{qw}).  By including
the conjugates of these, we have now exhausted all the possibilities
of the transitions that can happen following the connectivity of
vertex $m$. These rules can be applied to vertices with any number of
connections, and serve as the basis for the correspondence of the
graph to the QSW.

To complete the generalization, we need an equation of motion for a
quantum stochastic process. We choose the Kossakowski-Lindblad master
equation \cite{Kossakowski72a,Lindblad76a,Gorini76a},
$\frac{d}{dt}\rho= \mathcal{L}\big[ \rho \big]=\sum_k-i\left[H,\rho
\right]- \frac{1}{2}L_k^\dagger L_k\rho-\frac{1}{2}\rho L_k^\dagger
L_k+L_k \rho L_k^\dagger$, which evolves the density matrix as a
quantum stochastic process inspired by environmental effects under the
Markov approximation.  The Hamiltonian term describes the coherent
evolution (Schr\"odinger equation) while the rest describe a
stochastic evolution. We now set
$\mathcal{M}\rightarrow \mathcal{L}$ in Eq. (\ref{qsw}), where,
\begin{align}\label{kossak}
\mathcal{L}^{a\alpha}_{b\beta}
&= \sum_k\delta_{\alpha\beta} \langle a \vert \left( -iH-\frac{1}{2}L_k^\dagger L_k\right) \vert b \rangle \\
&+\delta_{ab} \langle \beta \vert \left( iH-\frac{1}{2}L_k^\dagger L_k\right) \vert \alpha \rangle
+ \langle a \vert L_k \vert b \rangle \langle \beta \vert L_k^\dagger \vert \alpha \rangle. \nonumber
\end{align}
\begin{table*}
\begin{tabular}{c c l} 
\hline\hline                        
Axiom & \;Matrix elements' connectivity &  \qquad Transition rate \\ [0.5ex]
\hline \hline                    
1 &  $\vert m \rangle \langle m \vert \leftrightarrows \vert m \rangle \langle m \vert$  &  $\mathcal{L}^{mm}_{mm}=\;\langle m \vert L_k\vert m \rangle \langle m \vert L_k^\dagger \vert m \rangle -\langle m \vert L_k^\dagger L_k\vert m \rangle$ \\ 
2 &  $\vert m \rangle \langle m \vert \leftrightarrows \vert n \rangle \langle n \vert$ & $\mathcal{L}_{mm}^{nn}=\; \langle n \vert L_k\vert m \rangle \langle m \vert L_k^\dagger \vert n \rangle $ \\ [1ex] 
\hline 
3 & $\vert m \rangle \langle m \vert \leftrightarrows \vert m \rangle \langle n \vert$ & $\mathcal{L}^{mn}_{mm}=\; \langle m \vert L_k\vert m \rangle  \langle m \vert L_k^\dagger \vert n \rangle + i\langle m \vert H\vert n \rangle-\frac{1}{2}\langle m \vert L_k^\dagger L_k\vert n \rangle$ \\ 
4  & $\vert m \rangle \langle n \vert \leftrightarrows \vert m \rangle \langle n \vert $ & $\mathcal{L}^{mn}_{mn}=\;\langle m \vert L_k\vert m \rangle \langle n \vert L_k^\dagger \vert n \rangle-i\langle m \vert H\vert m \rangle+i\langle n \vert H\vert n \rangle-\frac{1}{2}\langle m \vert L_k^\dagger L_k\vert m \rangle -\frac{1}{2}\langle n \vert L_k^\dagger L_k\vert n \rangle$ \\
5  & $\vert m \rangle \langle n \vert \leftrightarrows \vert l \rangle \langle n \vert $& $\mathcal{L}^{ln}_{mn}=\; \langle l \vert L_k\vert m \rangle \langle n\vert L_k^\dagger \vert n \rangle-i\langle l \vert H\vert m \rangle-\frac{1}{2}\langle l \vert L_k^\dagger L_k\vert m \rangle$\\ [1ex]
\hline 
6  & $\vert m \rangle \langle m \vert \leftrightarrows \vert l \rangle \langle n \vert$ & $\mathcal{L}^{ln}_{mm}=\; \langle l \vert L_k\vert m \rangle  \langle m \vert L_k^\dagger \vert n \rangle$ \\ 
[1ex]       
\hline \hline     
\end{tabular}  
\caption{Axioms for the quantum stochastic walk for processes that connect vertex $\ket{m}$ to its neighbors. Sum over $k$, and conjugate elements are implied. Axioms (1) and (2) correspond to the classical random walk. Axioms (3), (4) and (5) contain the quantum walk using $H$ and additional evolution due to $\{L_k\}$ terms. Axiom (6) comes only from the quantum stochastic walk, having no equivalent in the classical random walk or quantum walk.} 
\label{connectivity}   
\end{table*}

The connectivity conditions between a vertex $m$ and some connected
vertices $n$ and $l$ and the corresponding non-zero transition rates
according to Eq.~(\ref{kossak}) can be summarized in Table
\ref{connectivity} \footnote{For compactness, we assume that all connections
are equally weighted; the general case can be obtained by having
coefficients that depend on the vertices.}.

The Axioms from Table \ref{connectivity} capture the behavior sketched
in Fig.~(\ref{venn}). To recover the CRW, it suffices to consider
Axioms (1) and (2). These Axioms are a classical subset of the
transition rates of the QSW. On the other hand, the QW is obtained by
making all the rates of each element of $\{L_k\}$ zero. In this case,
only the subset of Axioms (3), (4) and (5) from Table
\ref{connectivity} are relevant, corresponding to the Hamiltonian, and
the QW is recovered. If the rates of $\{L_k\}$ are nonzero, these
Axioms contain behavior beyond the QW. Finally, Axiom (6) has
properties that have no equivalent in either the CRW or the QW; it is
a type of transition that appears exclusively in the QSW and leads to
different distributions. The choice of $H$ or $\{L_k\}$ is not
uniquely determined by the Axioms.

For example, a CRW is equivalent to a QSW with the choice of each
element of $\{L_k\}$ to be associated to $M$ in the following
manner. First, since connections are defined between two vertices, it
is easier to write the index $k$ in terms of two indices
$k\equiv(\kappa,\kappa^\prime)$. Using this notation, we enforce the graph by choosing
$L_{(\kappa,\kappa^\prime)}=M_\kappa^{\kappa^\prime}\vert \kappa
\rangle\langle \kappa^\prime\vert$. The Hamiltonian rates of $H$ are
set to zero. This choice ensures all the transition rates to be zero
except Axiom (2) from Table \ref{connectivity}, thereby recovering the
CRW.  An interpolation between the CRW and the QW conditions can be
used to obtain the classical-to-quantum transition of any graph by
changing the relative rates of the unitary and non-unitary processes.

\begin{figure}[htb]
\begin{center}
\resizebox{\columnwidth}{.7\columnwidth}{\includegraphics{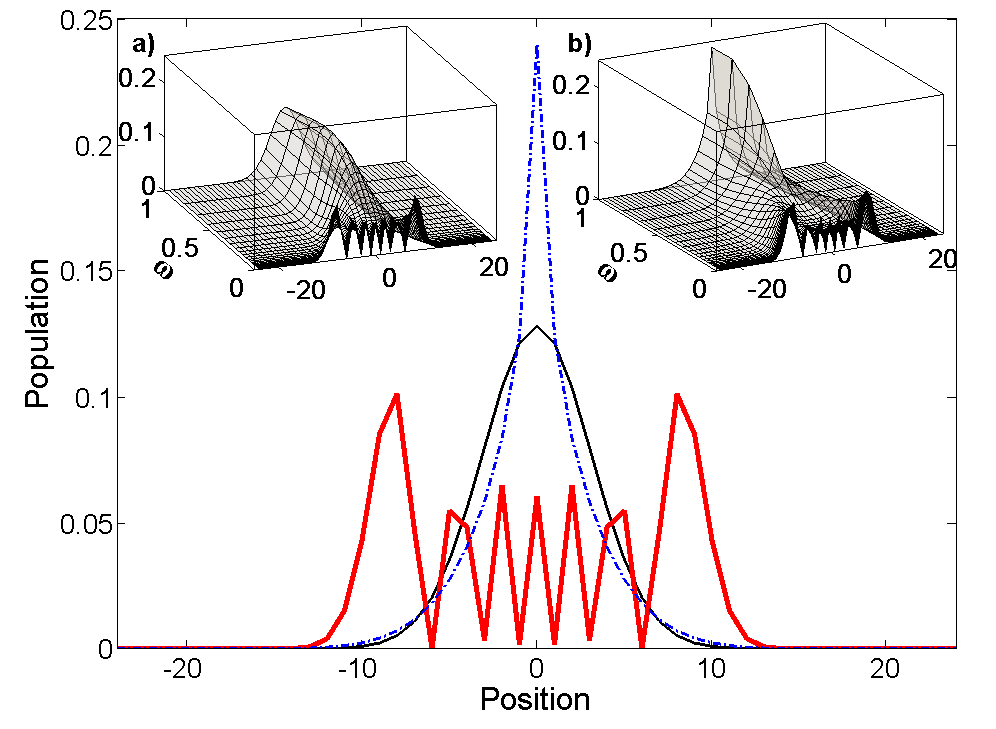}}
\end{center}  
\caption{Probability distributions of a walker starting at position $0$ of an infinite line defined by $M$ are shown at time $t=5$.  The heavy line represents the quantum walk, the thin line represents the classical random walk and the dotted line represents a choice of a quantum stochastic walk  due to a single operator $L=\sum_{\kappa,\kappa^\prime}M_\kappa^{\kappa^\prime}\vert  \kappa \rangle\langle \kappa^\prime\vert$ whose distribution cannot be obtained from the classical or quantum walks. \textbf{a)} The transition of the quantum walk to the classical random walk is shown.  The diagonal elements of the density matrix (Population) are plotted as a function of the vertex of the walker (Position) and a coupling parameter $\omega$.  When $\omega=0$ the evolution is due only to the Hamiltonian, and when $\omega=1$ the Hamiltonian dynamics are not present with the environmental dynamics taking over.  \textbf{b)} The transition of the quantum walk to the quantum stochastic walk is shown.}
\label{fig1}
\end{figure}

Other sets of $\{L_k\}$ can yield behavior different from both the
classical and quantum walks.  To illustrate the difference between
these choices, we study a simple example: the walk
on an infinite line, where a vertex $j$ is only connected to
its nearest neighbors. The conditions for the dynamics of the walk on
the line can be obtained from Table~\ref{connectivity} by making
$j=m$, $j-1=l$ and $j+1=n$. To obtain behavior that is different from
both the CRW and the QW, we specify a set $\{L_k\}$ with only one
member that is
$L=\sum_{\kappa,\kappa^\prime}M_\kappa^{\kappa^\prime}\vert \kappa
\rangle\langle \kappa^\prime\vert$. This choice makes Axiom (6)
nonzero, a type of transition that cannot be interpreted either as the
CRW or the QW. We illustrate the resulting distribution, and compare
it to CRW and the QW, in Fig.~\ref{fig1}.

The QSW provides a general framework to study the transition between
the CRW and the QW. In Fig. \ref{fig1}a, we use the parameter
$\omega=[0,1]$ to interpolate between the QW and the CRW:
$\mathcal{L}_\omega \left[ \rho\right]=-(1-\omega)\,i\left[ H,\rho
\right]+\omega\sum_k\left(- \frac{1}{2}L_k^\dagger
  L_k\rho-\frac{1}{2}\rho L_k^\dagger L_k+L_k \rho
  L_k^\dagger\right)$. By combining the unitary evolution and the
evolution due to the $\{L_{(\kappa,\kappa^\prime)}\}$ terms, the transition
from the QW to the CRW can be studied as one quantum stochastic
walk. This procedure can be done in general to study the transition
from any QW to a CRW for any graph \footnote{Our decoherence
model comes from the graph, while the ones discussed in
Ref. \cite{Kendon07a} were inspired by a physical model that localizes
the position and thus give different distributions. We propose studies
like Ref. \cite{Kendon07a} could be reinterpreted as a QSW, where the
environmental part, as interpreted from Table \ref{connectivity}, affects the connectivity of the graph
and thus generates
different distributions.}. To highlight the difference between the QSW
and the QW, we show the transition between them with
$L=\sum_{\kappa,\kappa^\prime}M_\kappa^{\kappa^\prime}\vert \kappa
\rangle\langle \kappa^\prime\vert$ on Fig.~\ref{fig1}b.

Although in this paper we focused on continuous-time walks, a parallel argument holds for discrete-time walks.  A discrete-time CRW evolves each time step by a Markov chain with a stochastic matrix $\mathcal{S}$ following $p_a=\sum_b \mathcal{S}_b^a\,p_b$. The quantum analogue is to use a quantum stochastic map $\mathcal{B}$ \cite{Sudarshan61a} by means of $\rho^\prime_{a\alpha}=\sum_{b\beta} \mathcal{B}_{b\beta}^{a\alpha}\,\rho_{b\beta}$, or equivalently as a superoperator, $\rho^\prime=\mathcal{B}\left[\rho\right]\equiv\sum_k\mathcal{C}_k\rho\mathcal{C}^\dagger_k$. A similar set of Axioms to Table \ref{connectivity} can be computed by using $ \mathcal{B}^{a\alpha}_{b\beta}=\sum_k\langle a \vert\mathcal{C}_k\vert b \rangle \langle \beta \vert\mathcal{C}^\dagger_k\vert \alpha \rangle$ \footnote{There is no need for an explicit coin, as the stochastic map is the most general evolution of a density matrix, including random processes. However, if desired, a coin could be implemented as an environment from the open quantum system interpretation of the map.}. The connection between the discrete-time QW and the continuous-time QW has been studied by Strauch \cite{Strauch06a}.

In conclusion, we introduced an axiomatic approach to define the
quantum stochastic walk, which behaves according
to a quantum stochastic process as constrained by a particular graph. This walk recovers the classical random walk and the quantum
walk as special cases, as well as the transition between them. The quantum stochastic walk
allows for the construction of new types of walks that result in
different probability distributions. As an example, we studied the
walk on a line.  The quantum stochastic walk provides a framework to
study quantum walks under decoherence. Reexamination of previous
work that considered environmental effects from physical motivations
might suggest that, if interpreted as a quantum stochastic walk, the
environment is changing the effective graph. Since the quantum stochastic walk is more general than
the quantum walk, it is therefore universal for quantum computation. Its quantum to classical transiton can also be used to examine decoherence phenomena
in quantum computation and classical-to-quantum phase
transitions. New quantum stochastic distributions might suggest new kinds of quantum algorithms
or even new classes of quantum algorithms based on this model.

{\bf Acknowledgments}: We thank F. Strauch for many insightful
discussions and comments. We also thank I. Kassal, L. Vogt, S. Venegas-Andraca, P. Rebentrost, K. Modi,
S. Jordan and E.C.G. Sudarshan for their comments on the
manuscript. A.A.G. and J.D.W. thank the Army Research Office under
contract W911NF-07-1-0304 and C.A.R. thanks the Mary-Fieser
Postdoctoral Fellowship program. 

 \bibliography{refs2008}

\end{document}